\documentclass[conference]{IEEEtran}
\usepackage[nolist]{acronym}

\begin{acronym}[ACC]
\acro{ACC}{Adaptive Cruise Control}
\end{acronym}
\begin{acronym}[ADS]
\acro{ADS}{automated driving system}
\end{acronym}
\begin{acronym}[AoI]
\acro{AoI}{Area of Interest}
\end{acronym}
\begin{acronym}[ARQ]
\acro{ARQ}{Automatic Repeat Request}
\end{acronym}
\begin{acronym}[CACC]
\acro{CACC}{Cooperative Adaptive Cruise Control}
\end{acronym}
\begin{acronym}[CAM]
\acro{CAM}{Cooperative Awareness Message}
\end{acronym}
\begin{acronym}[CBR]
\acro{CBR}{Channel Busy Ratio}
\end{acronym}
\begin{acronym}[CA]
\acro{CA}{Cooperative Awareness}
\end{acronym}
\begin{acronym}[CBF]
\acro{CBF}{Contention-Based Forwarding}
\end{acronym}
\begin{acronym}[CCAM]
\acro{CCAM}{Cooperative, Connected and Automated Mobility}
\end{acronym}
\begin{acronym}[C-ITS]
\acro{C-ITS}{Cooperative Intelligent Transport Systems}
\end{acronym}
\begin{acronym}[CP]
\acro{CP}{Collective Perception}
\end{acronym}
\begin{acronym}[CPM]
\acro{CPM}{Collective Perception Message}
\end{acronym}
\begin{acronym}[DCC]
\acro{DCC}{Decentralized Congestion Control}
\end{acronym}
\begin{acronym}[DDT]
\acro{DDT}{Dynamic Driving Task}
\end{acronym}
\begin{acronym}[DEN]
\acro{DEN}{Decentralized Environmental Notification}
\end{acronym}
\begin{acronym}[DENM]
\acro{DENM}{Decentralized Environmental Notification Message}
\end{acronym}
\begin{acronym}[DPD]
\acro{DPD}{Duplicate Packet Detection}
\end{acronym}
\begin{acronym}[DPL]
\acro{DPL}{Duplicate Packet List}
\end{acronym}
\begin{acronym}[EDCA]
\acro{EDCA}{Enhanced Distributed Channel Access}
\end{acronym}
\begin{acronym}[ETSI]
\acro{ETSI}{European Telecommunications Standards Institute}
\end{acronym}
\begin{acronym}[EPAC]
\acro{EPAC}{Electrically Power Assisted Cycle}
\end{acronym}
\begin{acronym}[FCD]
\acro{FCD}{Floating Car Data}
\end{acronym}
\begin{acronym}[IGG]
\acro{IGG}{Inter-generation Gap}
\end{acronym}
\begin{acronym}[IPG]
\acro{IPG}{Inter-packet Gap}
\end{acronym}
\begin{acronym}[ITS]
\acro{ITS}{Intelligent Transport Systems}
\end{acronym}
\begin{acronym}[ITS-S]
\acro{ITS-S}{ITS Station}
\end{acronym}
\begin{acronym}[LocT]
\acro{LocT}{Location Table}
\end{acronym}
\begin{acronym}[LocTE]
\acro{LocTE}{Location Table entry}
\end{acronym}
\begin{acronym}[MC]
\acro{MC}{Maneuver Coordination}
\end{acronym}
\begin{acronym}[MCO]
\acro{MCO}{Multi-channel Operation}
\end{acronym}
\begin{acronym}[MCM]
\acro{MCM}{Maneuver Coordination Message}
\end{acronym}
\begin{acronym}[mmWave]
\acro{mmWave}{millimiter wave}
\end{acronym}
\begin{acronym}[OBU]
\acro{OBU}{On-board Unit}
\end{acronym}
\begin{acronym}[OEDR]
\acro{OEDR}{Object and Event Detection and Reaction}
\end{acronym}
\begin{acronym}[PAM]
\acro{PAM}{Platooning Awareness Message}
\end{acronym}
\begin{acronym}[PCM]
\acro{PCM}{Platooning Control Message}
\end{acronym}
\begin{acronym}[P2X]
\acro{P2X}{Pedestrian-to-Anything}
\end{acronym}
\begin{acronym}[PDR]
\acro{PDR}{Packet-delivery Ratio}
\end{acronym}
\begin{acronym}[RadCom]
\acro{RadCom}{Radar-Based Communication}
\end{acronym}
\begin{acronym}[RSU]
\acro{RSU}{Road-side Unit}
\end{acronym}
\begin{acronym}[RHW]
\acro{RHW}{Road Hazard Warning}
\end{acronym}
\begin{acronym}[SHB]
\acro{SHB}{Single-Hop Broadcasting}
\end{acronym}
\begin{acronym}[S-CBR]
\acro{S-CBR}{Smoothed Channel Busy Ratio}
\end{acronym}
\begin{acronym}[TC]
\acro{TC}{Traffic Class}
\end{acronym}
\begin{acronym}[TDC]
\acro{TDC}{Transmit Data Rate Control}
\end{acronym}
\begin{acronym}[TLS]
\acro{TLS}{Transport Layer Security}
\end{acronym}
\begin{acronym}[TPC]
\acro{TPC}{Transmit Power Control}
\end{acronym}
\begin{acronym}[TRC]
\acro{TRC}{Transmit Rate Control}
\end{acronym}
\begin{acronym}[TTL]
\acro{TTL}{Time-to-Live}
\end{acronym}
\begin{acronym}[VBS]
\acro{VBS}{Vulnerable Road User Awareness basic service}
\end{acronym}
\begin{acronym}[VAM]
\acro{VAM}{VRU Awareness Message}
\end{acronym}
\begin{acronym}[VANET]
\acro{VANET}{Vehicular ad hoc Network}
\end{acronym}
\begin{acronym}[VRU]
\acro{VRU}{Vulnerable Road User}
\end{acronym}
\begin{acronym}[V2X]
\acro{V2X}{Vehicle-to-Anything}
\end{acronym}
\begin{acronym}[V2V]
\acro{V2V}{Vehicle-to-Vehicle}
\end{acronym}
\IEEEoverridecommandlockouts

\usepackage{cite}
\usepackage{amsmath,amssymb,amsfonts}
\usepackage{algorithmic}
\usepackage{booktabs}
\usepackage{graphicx}
\usepackage{textcomp}
\usepackage{xcolor}
\def\BibTeX{{\rm B\kern-.05em{\sc i\kern-.025em b}\kern-.08em
    T\kern-.1667em\lower.7ex\hbox{E}\kern-.125emX}}
\begin{document}

\title{	Non-Negotiated Implicit ETSI VAM Clustering
\thanks{This work was sponsored by the Swedish Transport administration through the Skyltfonden project 2023/104170, and by the ELLIIT network through project 6G-V2X "6G Wireless, sub-project: Vehicular Communications".}
}

\author{\IEEEauthorblockN{1\textsuperscript{st} Felipe Valle}
\IEEEauthorblockA{\textit{School of Information Technology} \\
\textit{Halmstad University}\\
Halmstad, Sweden \\
felipe.valle@hh.se}
\and\IEEEauthorblockN{2\textsuperscript{nd} Daniel Bleckert}
\IEEEauthorblockA{\textit{School of Information Technology} \\
\textit{Halmstad University}\\
Halmstad, Sweden \\
danble21@student.hh.se}
\and
\IEEEauthorblockN{3\textsuperscript{rd} Linus Frisk}
\IEEEauthorblockA{\textit{School of Information Technology} \\
\textit{Halmstad University}\\
Halmstad, Sweden \\
linufr21@student.hh.se}
\and
\IEEEauthorblockN{4\textsuperscript{th} Oscar Amador Molina}
\IEEEauthorblockA{\textit{School of Information Technology} \\
\textit{Halmstad University}\\
Halmstad, Sweden \\
oscar.molina@hh.se}
\and
\IEEEauthorblockN{5\textsuperscript{th} Elena Haller}
\IEEEauthorblockA{\textit{School of Information Technology} \\
\textit{Halmstad University}\\
Halmstad, Sweden \\
elena.haller@hh.se}
\and
\IEEEauthorblockN{6\textsuperscript{th} Alexey Vinel}
\IEEEauthorblockA{
\textit{Karlsruhe Institute of Technology}\\
Karlsruhe, Germany \\
alexey.vinel@kit.edu}
}

\maketitle

\begin{abstract} Including \ac{VRU} in \ac{C-ITS} framework aims to increase road safety. However, this approach implies a massive increase of network nodes and thus is vulnerable to medium capacity issues, e.g., contention, congestion, resource scheduling. Implementing cluster schemes ---to reduce the number of nodes but represent the same number of \acp{VRU}--- is a direct way to address the issue. One of them is suggested by \ac{ETSI} and consists of nodes (connected pedestrians and cyclists) sending vicarious messages to enable a leader node to cover for a cluster of \acp{VRU}. However, the proposed scheme includes negotiation to \textit{establish} a cluster, and \textit{in-cluster} communication to maintain it, requiring extra messages of variable sizes and thus does not fully resolve the original medium capacity issues. Furthermore, these exchanges assume network reliability (i.e. a lossless channel and low latency to meet time constraints).  We propose a method for \ac{VAM} clustering where 1) all cluster operations are performed without negotiation, 2) cluster leaders do not require sending additional messages or meet deadlines, and 3) assumes a lossy communication channel and offers a mechanism for cluster resilience. Our results show the feasibility of the concept by halving message generations compared to individual messages while keeping the awareness levels (i.e., that VRUs are accounted for).
\end{abstract}


\section{Introduction}
The safety of \acp{VRU} --- e.g., cyclists, pedestrians --- is one of the purposes of \ac{C-ITS} and \ac{V2X} technologies as a part of \ac{CCAM} is a way \acp{C-ITS} follow to protect \acp{VRU}. Standardization bodies such as \ac{ETSI} have defined services enabling \acp{VRU} share their current and future (via heading) positions. This is realized by means of  \ac{VBS}~\cite{etsi-va} --- vehicles receive \acp{VAM} to become aware of the presence of \acp{VRU}.

However, \ac{V2X} messages do not always reach their destination due to congestion~\cite{Amador2020} or contention~\cite{BaiocchiAoI} problems. This was taken into account when designing the ETSI \ac{VBS}, which considers the possibility of grouping \acp{VRU} into clusters~\cite{etsi-va}. In a cluster of \acp{VRU} with homogeneous behavior, a leader sends vicarious messages that cover for an area between 3--5\,m and 3--20 \acp{VRU}. While cluster members stop sending \acp{VAM}, they still have to perform in-cluster communication, i.e. to report joining, leaving, shrinking or enlarging either in the 5.9\,GHz \ac{V2X} medium or \textit{out of band} (e.g., using Bluetooth). Furthermore, the specification goes as far as requiring a \ac{VRU} to indicate the reason for leaving a cluster~\cite[5.4.2]{etsi-va}.

This work presents an alternative method for clustering \acp{VAM} such that
\begin{itemize}
    \item[-] neither negotiations for cluster operations are required
    \item[-] nor communications for cluster operations occur out of band (i.e., our \acp{VRU} are only equipped with \ac{V2X} nodes), and
    \item[-] assumes a lossy communication channel but offers a mechanism for cluster resilience.
\end{itemize}
 To evaluate the performance, the expected medium usage for individual \acp{VAM} generation (i.e., \textit{standalone}) is taken as a baseline.
 
 The rest of the document is organized as follows: Section~\ref{sec:background} presents the technical framework for VAM clustering, the new clustering scheme is proposed in Section~\ref{sec:proposal} and evaluated in Section~\ref{sec:evaluation} through simulation. A discussion and conclusions are presented in Sections~\ref{sec:discussion} and \ref{sec:conclusion}, respectively.

\section{Background}
\label{sec:background}

\subsection{ETSI VRU Awareness Messages}
\begin{figure}[t!]
    \centering
    \includegraphics[width=1\linewidth]{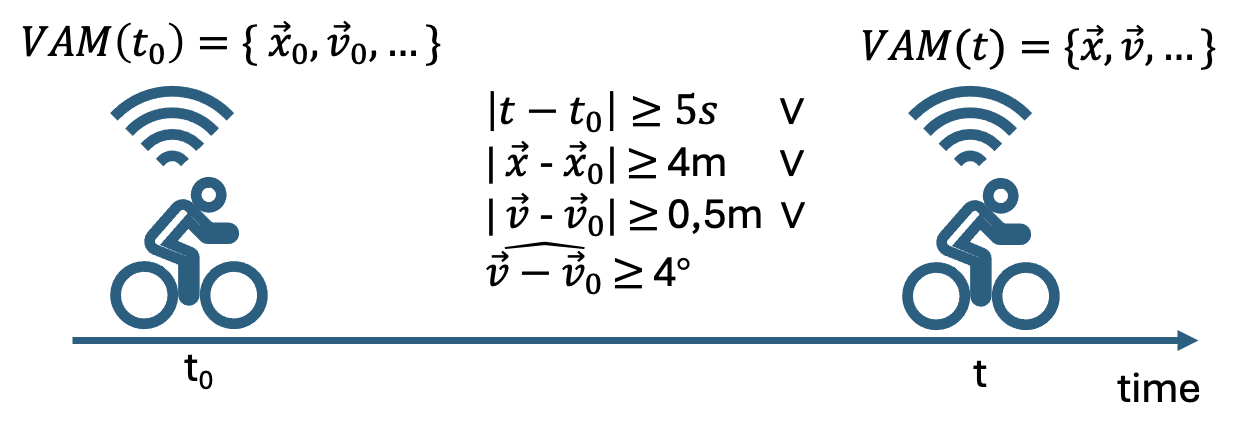}
    \caption{VAM generation conditions}
    \label{fig:vam}
\end{figure}

\ac{ETSI} has defined a framework for ITS where road users send messages that power applications for road safety and traffic efficiency. Services such as the \ac{CA}~\cite{etsiCAM}, \ac{DEN}~\cite{etsiDENM} basic services, and \ac{VBS} generate messages that are event-driven (e.g., \acp{DENM} that alert of dangerous events on the road), or semi-periodical (e.g., \acp{CAM} and \acp{VAM} that inform about road users' location and heading). Fig.\ref{fig:vam} summarizes how \acp{VAM} are triggered by changes in position, speed or heading. Every 0.1\,s, a \ac{VRU} checks if it had exceeded thresholds. If any of these conditions is met, it generates a \ac{VAM} that is then sent down the layers to be transmitted using any of the Access layer technologies available for the node (e.g., ETSI ITS-G5, LTE, 5G-NR). Other connected road users then receive the \ac{VAM} and consume its information (e.g., they update their local dynamic map).

\subsection{ETSI VAM Clustering}

\begin{figure}[t!]
    \centering
    \includegraphics[width=1\linewidth]{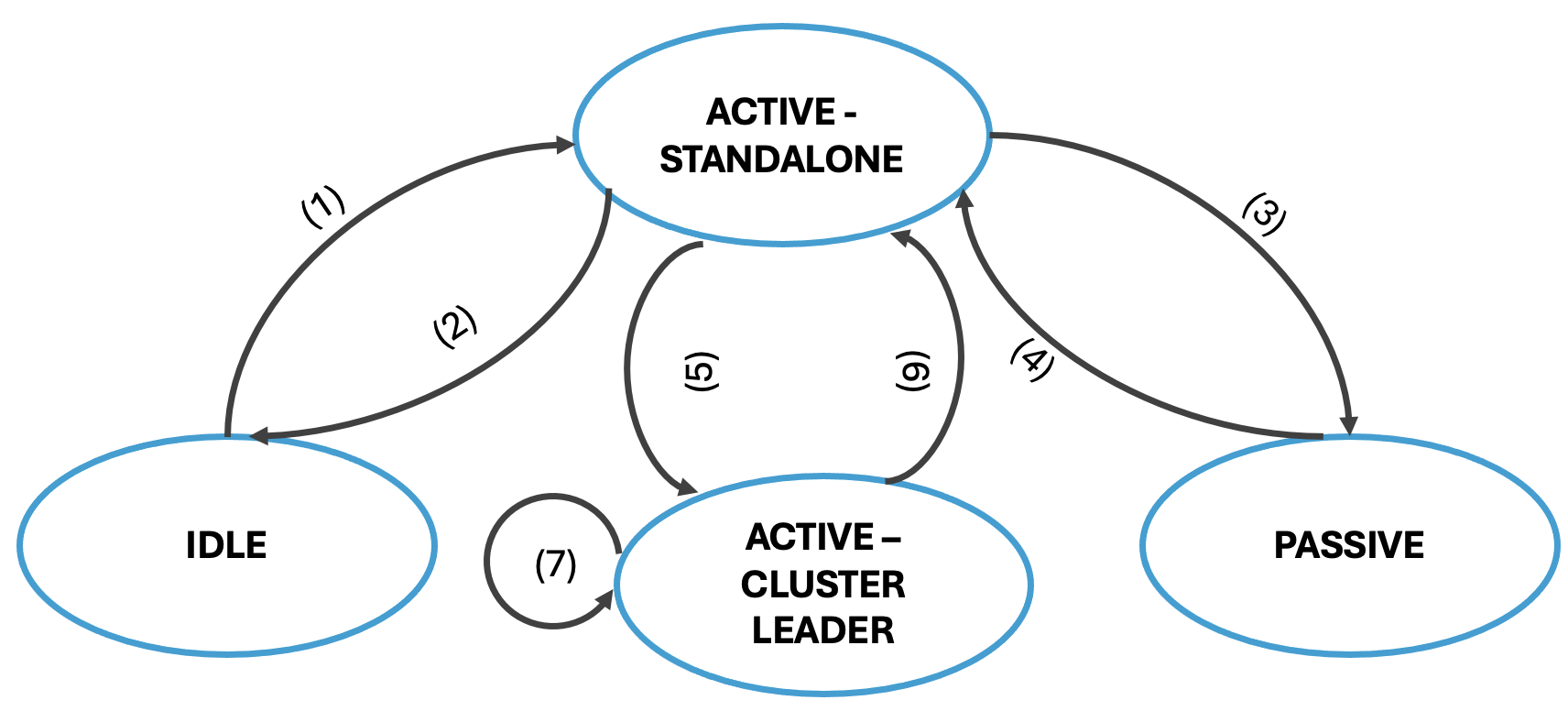}
    \caption{State diagram for ETSI VAM cluster operations}
    \label{fig:state_etsi}
\end{figure}

\begin{table}[t!]
    \centering
    \caption{State Transitions - ETSI VAM clustering}
    \label{tab:etsi_transitions}
    \begin{tabular}{p{1cm} p{5cm}}
    \toprule
       Label & Description \\
        \midrule
        (1) & ego starts transmitting VAMs\\
        (2) & ego stops transmitting VAMs\\
        (3) & the cluster has not reached its maximum number of members (e.g., 20), ego is at a certain distance away from the cluster leader (e.g., 3--5\,m), and velocity of the leader is within 5\% \cite{etsi-va}.\\ 
        (4) & ego detects cluster leader as lost or cluster conditions are not satisfied\\
        (5) & no cluster to join, ego has sufficient processing power, and it receives messages from 3--5 different \acp{VRU} located within 3--5\,m. \\
        (6) & ego breaks up the cluster it created (not enough power, no neighbors within a range)\\
        (7) & shrinking/extending cluster, ''keep alive''-messages (at least every 2 s)\\
        \bottomrule
    \end{tabular}
    
\end{table}

\begin{table}[b]
	\centering
	\caption{VRU States}
	\label{tbl:states}
	\begin{tabular}{ c  c  c}
		\toprule
		\textbf{State}  & \textbf{Runs VBS} & \textbf{Transmits VAMs} \\
		\midrule
		Idle & $\times$ & $\times$ \\
        Active - standalone & \checkmark & \checkmark \\
        Active - cluster leader &  \checkmark & \checkmark \\
        Passive &  \checkmark & $\times$\dag \\
        \midrule
		Active - cluster member &  \checkmark & \checkmark \ddag \\
		\bottomrule
	\end{tabular}\\
    \raggedright
    \dag Only to leave the cluster and move to Active - standalone\\
    \ddag If no message is received from leader after a timeout
\end{table}

Let us assume that a \ac{VRU} (\textit{ego}) is equipped with an \ac{ITS-S} that can send and receive \acp{VAM}. Fig.~\ref{fig:state_etsi} shows the different states an \ac{ITS-S} can be in. Active or Passive mean that a station generates \acp{VAM} or not, respectively. Table~\ref{tab:etsi_transitions} shows a summary of transitions between the states.

Communications between cluster members for cluster operations can occur either \textit{in-band} (i.e., using the 5.9\,GHz spectrum) and \textit{out-of-band} (e.g., using Bluetooth or ultra-wideband). These communications are used to maintain, leave, or break-up a cluster. A member can decide to leave the cluster if it does not meet the kinematic conditions or if it it leaves its role as a VRU (e.g., if it enters a bus). If the member leaving the cluster is the leader, then the cluster is broken up and members have to move to standalone mode.

Cluster data is sent by the leader and members in a specific container in a \ac{VAM}. Some of the data in the container is: an identifier for the cluster, a cluster shape (e.g., rectangular, circular), and data to indicate operations like joining, leaving, and breaking up. Thus, cluster containers add to the variability of sizes between \acp{VAM}, which has been identified as a cause for reduced network performance~\cite{Anupama2022} and thus, has an impact on safety metrics (e.g., colliding messages make road users blind to each other). These collisions are particularly damaging when a VAM from the leader is lost, since that message covers for multiple VRUs and members cannot make up for that message until they determine the leader has been lost and go back to standalone mode~\cite{sjostrom2023vam}.

The specification for the ETSI \ac{VBS} has a set of cluster membership parameters where some time periods are defined for cluster operations \cite[Table 15]{etsi-va}. The cluster head is required to send messages at least every 2 seconds to meet the \textit{timeClusterContinuity} parameter (after which members leave the cluster). Previous works (\cite{olsson2023intelligent,martin2023towards}) show that pedestrians walking at 5\,km/h generate messages at lower rates (closer to 0.333\,Hz), which implies that having them as cluster leaders will prompt them to generate more messages and potentially bring up contention issues at the Access layer. Furthermore, this 0.333\,Hz mark coincides with other timers set in the specification, e.g., \textit{timeClusterJoinNotification}, which is the message sent to request the leader to include it in its cluster, is set to 3\,s --- thus, it might only be sent in one VAM. Finally, even if the message requesting to join a cluster is received, the leader has 0.5\,s to notify (through an updated cluster VAM) that the member has joined successfully (\textit{timeClusterJoinSuccess}). Thus, the leader might be forced to send confirmation messages (i.e., a cluster VAM including the new member) at even higher rates, once again, bringing up potential problems at the Access layer.

\section{Implicit VAM Clustering}
\label{sec:proposal}

\begin{figure}[t!]
    \centering
    \includegraphics[width=1\linewidth]{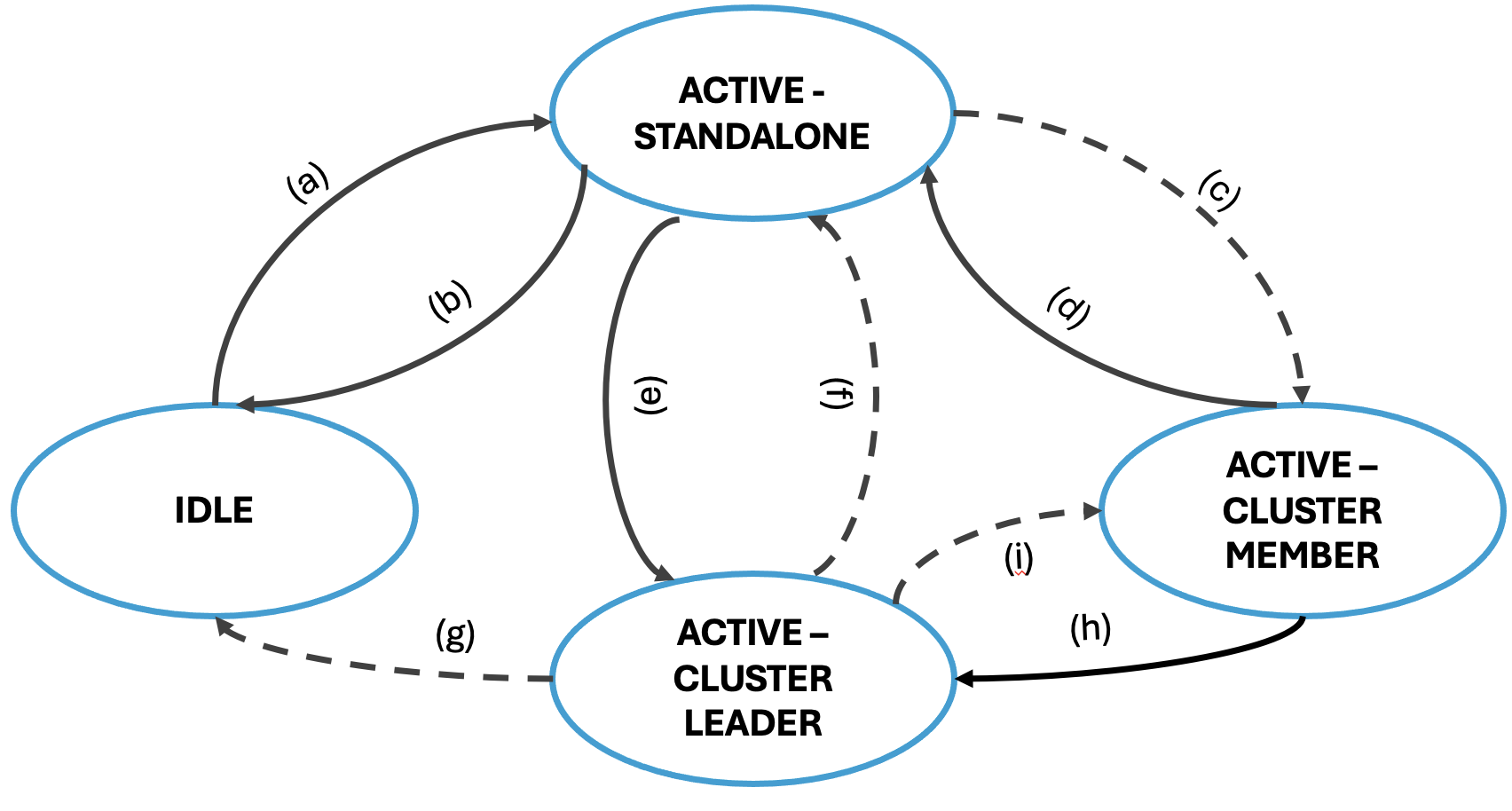}
    \caption{State diagram for Implicit Clustering}
    \label{fig:state_implicit}
\end{figure}

\begin{table}[t!]
    \centering
    \caption{State Transitions - Implicit clustering}
    \label{tab:implicit_transitions}
    \begin{tabular}{p{1cm} p{5cm}}
    \toprule
       Label & Description \\
        \midrule
        (a) & ego starts transmitting VAMs\\
        (b) & ego stops transmitting VAMs\\
        (c) & ego is within cluster range from the leader (cluster conditions are met)\\ 
        (d) & cluster conditions are not satisfied \\
        (e) & no cluster to join, ego has sufficient processing power, and it receives messages from \acp{VRU} located within 5\,m. \\
        (f) & ego breaks up the cluster it created (stop offering coverage)\\
        (g) & ego stops transmitting VAMs\\
        (h) & leader did not send a VAM when expected plus a random jitter and ego sends a VAM offering coverage\\
        (i) & ego receives a message from a close VRU offering coverage and yields leadership\\
        \bottomrule
    \end{tabular}
    
\end{table}

We propose a clustering scheme where clustering operations do not require special containers, and banding or disbanding a cluster is done without negotiation (i.e., additional messaging). Fig.~\ref{fig:state_implicit} describes the process: as with ETSI, when ego is in standalone mode and listens to \acp{VAM} from one or more neighbors within 5\,m, it modifies the data in its next \ac{VAM} to indicate that it can offer coverage to is neighbors. Upon receiving that \ac{VAM} from ego, neighbors assess if they are within coverage, and they enter the \textit{active-cluster member} state, where they restart all their VAM kinematic triggers (position, speed, heading) and update their last VAM timestamp to now ($t_0$). Neighbors in the \textit{active - cluster member} state still keep track of their kinematic parameters, but they do not generate a \ac{VAM} immediately after a trigger is set off. Instead, they wait for an additional random timeout between 0.1--5\,s (standard values for minimum and maximum inter-VAM generations) to give a chance for ego to send a new \ac{VAM}. If ego sends a message still keeping coverage and the distance condition is still met, neighbors restart their triggers, and the process is repeated. If the message from ego is not received, if coverage is not sufficient, or if the distance conditions are not met, neighbors can then go back to the standalone state (i.e., they generate the VAM after that timeout and keep generating VAMs until they exit the standalone state). Dotted lines indicate processes that do not require a VAM to be sent.

The disbanding process for the cluster is performed by the leader. If it stops broadcasting its shape as a circle of $r = 5m$, members will automatically detect coverage as non-existent and will go into standalone mode where, if needed, they can offer coverage for other neighbors. Furthermore, if the message from the leader is lost, one of the members (the one with the shortest random timeout) will send a message offering coverage. This will prompt former members that still meet the new conditions to inhibit themselves from sending a VAM, and those who are out of coverage can go into the standalone and then create their own cluster or join an existing one.

This method keeps track intrinsically of other conditions such as speed homogeneity, since leaders or members that move at significantly different speeds will get out of coverage naturally, and those that have similar dynamics will tend to stay clustered. Additionally, a cluster leader can enter the idle status without needing to break-up the cluster first. Finally, our approach does not require the use of additional containers or messages to perform cluster operations and maintenance, which aids to keeping VAM sizes constant.

\section{Evaluation}
\label{sec:evaluation}

\begin{table}[b]
	\centering
	\caption{VAM clustering messages}
	\label{tbl:messages}
	\begin{tabular}{ c  c  c}
		\toprule
		\textbf{Operation}  & \textbf{ETSI Clustering} & \textbf{Implicit Clustering} \\
		\midrule
		Offer to lead & \textcolor{blue}{$\checkmark$} & \textcolor{blue}{$\checkmark$}\dag \\
        Cluster Join Request*& \textcolor{red}{\checkmark}\ddag & \textcolor{blue}{$\times$} \\
        Join Success Notification** & \textcolor{red}{\checkmark}\ddag & \textcolor{blue}{$\times$} \\
        Keep alive message &  \textcolor{red}{\checkmark}\ddag & \textcolor{blue}{\checkmark}\dag \\
        Break-up &  \textcolor{red}{\checkmark}\ddag & \textcolor{blue}{$\times$}\dag \\
		\bottomrule
	\end{tabular}\\
    \raggedright
    \dag Uses a regular VAM transmitted at kinematic-based rate\\
    \ddag Uses a new message with special containers not triggered by kinematics\\
    * One per member\\
    ** Up to one per member
\end{table}

Table~\ref{tbl:messages} shows the messages that are required to start, maintain, and break-up a cluster following the ETSI specification or Implicit Clustering. Let us picture a scenario where \textit{ego} --- a pedestrian on a straight path that generates VAMs every 3\,s --- is aware of other five \acp{VRU} that are potential members for a cluster. In this scenario, we would like to keep the cluster alive for 30\,s and then break it up. Let us then compare the number of expected messages to be sent in each scheme. For ETSI Clustering, \textit{ego} sends the offer to lead, and then each member will reply with a VAM with the cluster container and the necessary bytes set to the appropriate values (up to 5 messages). Depending on the time at which those messages arrived, \textit{ego} replies to these requests with one or up to five messages (since it must respond within 500\,ms, so it can potentially reply to everyone in one message). Then, as a cluster leader, it has to send cluster VAMs at least every 2\,s. Finally, \textit{ego} breaks up the cluster with a VAM. In the 30\,s of this scenario, this cluster created between 18 and 22 messages and that is considering all of them to have been successful. This is a reduction from the 10$\times$6 VAMs that VRUs would have generated in standalone mode. Nevertheless, in the same time, Implicit Clustering would need the same 10 VAMs that \textit{ego} would have generated individually.

\subsection{Numerical Evaluation}

We use the model for IEEE~802.11p presented in~\cite{BaiocchiAoI} to quantify the benefit of clustering and compare our Implicit Clustering scheme to what is expected from ETSI Clustering. We consider a scenario where nodes can \textit{listen} to each other: a 50\,m segment with two sidewalks each with a capacity of four pedestrians. Two densities are considered --- one with pedestrians with a mean separation of 4\,m, and one with pedestrians separated by 2\,m (i.e., 100 and 200 in total in the segment, respectively). Clustering schemes group VRUs in sets of 12 and 20 for each density. The model outputs the \ac{PDR}, which is the rate between successfully received messages against the total number of attempts.

Table~\ref{tab:numerical} shows that we can expect losses of \acp{VAM} in standalone mode. The clustering schemes perform quite similarly, with only a few loses per thousands of \acp{VAM}. However, the result for standalone transmissions has worse implications for ETSI clustering than for Implicit Clustering. On the one hand, if standalone messages are lost, either offers to join or requests to join are also lost. Furthermore, the additional exchanges add to an already stressed channel. The result for ETSI clustering is then conditional to the result of standalone. On the other hand, all Implicit Clustering requires is one node to listen to a neighbor within range to start offering coverage, and then, each additional neighbor that meets the conditions will inhibit itself from transmitting. Thus, implicit clustering will eventually reach steady state. In layman terms, ETSI clustering expects \textit{people} to 1) be asked to be silent, 2) offering to be silent, and 3) be confirmed that they were asked to be silent --- all within hard deadlines and without mechanisms to recover misunderstood requests --- while implicit clustering allows people to be silent because someone is already speaking on their behalf.

\begin{table}[tb]
    \centering
    \caption{Numerical Evaluation of VAM schemes}
    \label{tab:numerical}
    \begin{tabular}{r c c}
    \toprule
       \textbf{Scheme} & \multicolumn{2}{c}{\textbf{PDR}} \\
         & \textbf{$d = 4\,m$} & \textbf{$d = 2\,m$} \\
        \midrule
        Standalone & 94.74\% & 66.77\% \\
        ETSI Clustering & 99.53\% & 98.39\%\\
        Implicit Clustering & 99.78\% & 98.90\% \\
        \bottomrule
    \end{tabular}
    
\end{table}

\subsection{Simulation}

\begin{table}[t]
	\centering
	\caption{Simulation Parameters}
	\label{tbl:simpars}
	\begin{tabular}{ r  l }
		\toprule
		\textbf{Parameter}  & \textbf{Values} \\
		\midrule
		Access Layer protocol & ITS-G5 (IEEE 802.11p) \\
        Channel bandwidth & 10\,\textrm{MHz} at 5.9\,\textrm{GHz} \\
        Data rate & 6\,\textrm{Mbit/s} \\
        Pedestrian Transmit power & 16\,\textrm{mW}\\
        Vehicle Transmit power & 20\,\textrm{mW}\\
		Path loss model & Two-Ray interference model \\
        VAM generation frequency & 0.2--10~\textrm{Hz} (ETSI VAM~\cite{etsi-va}) \\
        CAM generation frequency & 1--10~\textrm{Hz} (ETSI CAM~\cite{etsiCAM}) \\
        Max. pedestrian velocity &  5\,\textrm{km/h} \\
        Max. vehicle velocity &  60\,\textrm{km/h} \\
        Pedestrian density &  48\,ped. per 100\,m segment \\
        Vehicular density &  30\,veh/km per lane \\
        Cluster shape and size &  Circle of $r=5m$ \\
		\bottomrule
	\end{tabular}
\end{table}

We evaluate our proposed clustering scheme and compare it against nodes sending standalone messages by means of simulations. We use the Artery~\cite{Artery} simulation toolkit where we place pedestrians on two sidewalks along a 2\,km road segment. Both sidewalks have pedestrians walking on both directions at a target speed of 5\,km/h~\cite{avgspeed}. All pedestrians are equipped with \acp{ITS-S} that only transmit in the 5.9\,Ghz band using ETSI ITS-G5 (i.e., IEEE\,802.11p). Only pedestrian traffic is included. Measurements are taken for 60\,s after a 100\,s warm-up period. We perform five repetitions and average values with 95\% confidence intervals are shown unless noted differently. Table~\ref{tbl:simpars} shows the rest of the simulation parameters.

We keep statistics for network and safety metrics:
\begin{itemize}
    \item \textbf{\ac{IPG}:} the mean time $\Delta t^r$ between two consecutive successful message receptions from the same node $j$
    \[\Delta t^r[i]=t^r_{VAM_{\leftarrow j}}[i+1]-t^r_{VAM_{\leftarrow j}}[i]\]
    \item \textbf{\ac{IGG}:} the mean time $\Delta t^g$ between two message generations.
       \[\Delta t^g[i]=t^g_{VAM_{\rightarrow}}[i+1]-t^g_{VAM_{\rightarrow}}[i]\]
       The closer \ac{IGG} and \ac{IPG} are, the better network reliability.
    \item \textbf{VRU awareness:} the ratio between the number of \acp{VRU} visible in the network against those present in the scenario.
    \[Awareness=\frac{N \{VRU:\Delta t^{r,g}[i]<3s\}} {N_{tot} \{VRU\}}\]
    We count the number of nodes with at least one successful broadcast within a period of 3\,s and divide it by the number of VRU \acp{ITS-S} in the given scenario.
\end{itemize}

\subsection{Results}
\begin{figure}[tb!]
    \centering
    \includegraphics[width=1\linewidth]{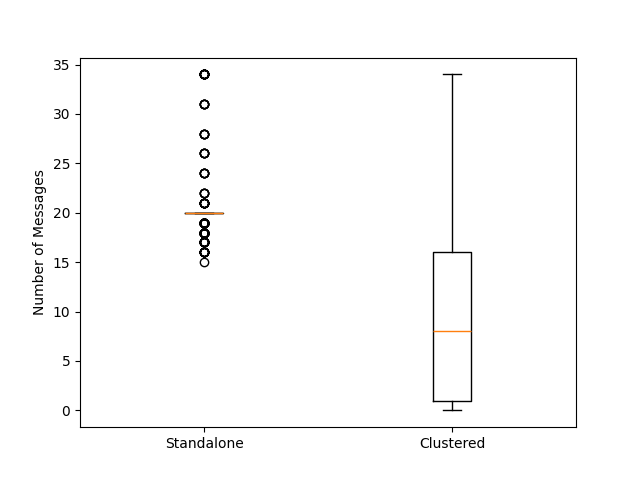}
    \caption{Generated VAMs per station}
    \label{fig:genvams}
\end{figure}

Fig~\ref{fig:genvams} shows the decrease in generations when clustering is applied. In standalone mode, \acp{VRU} trigger messages every 3\,s. The outliers represent those times in the simulation when a pedestrian is changing speed or orientation (e.g., when trying to pass another pedestrian). The result for our implicit clustering approach shows that there are nodes that never generate messages and the box plot shows that most nodes cut down generations, with implications, e.g., in power used for transmissions.

\begin{figure}
    \centering
    \includegraphics[width=1\linewidth]{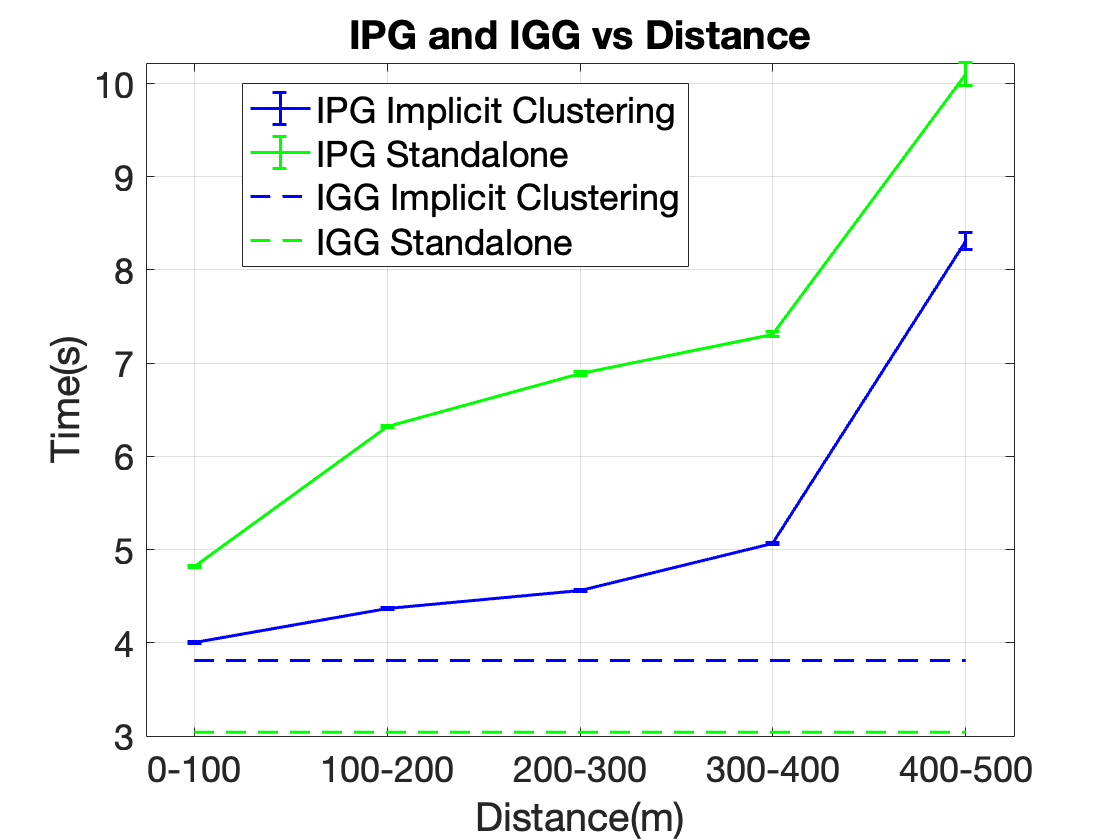}
    \caption{Inter-packet Gaps with VRU and vehicular traffic}
    \label{fig:ipg_cars}
\end{figure}

Fig.~\ref{fig:ipg_cars} shows a comparison of network metrics (i.e., IGG and IPG). In line with the results in Fig.~\ref{fig:genvams}, standalone nodes generate messages every 3\,s, while those that are clustered do so closer to every 4\,s. Network metrics, however, do not respond well to the greediness of the standalone mode, and successful receptions at short distances occur in gaps of above 4\,s for both schemes (with IPGs being closer to IGGs for the clustered mode). Propagation phenomena amplifies the shortcomings of the standalone mode, and there is a significant difference in IPGs when compared to the clustered approach, with the latter being constantly 2\,s better at medium to large distances. 

Even if CAM traffic brings medium congestion, \ac{DCC} does not seem to affect \ac{VAM} generation. \ac{DCC} limits rates between 1--40\,Hz, which is out of the expected rate for pedestrians and some bicycles (1/3--1\,Hz, respectively~\cite{olsson2023intelligent}). Thus, generation rates for \acp{VAM} cannot be controlled by \ac{DCC} --- shown by the dashed lines in Fig.~\ref{fig:ipg_cars}. There is a significant gap between IGG and IPG for the standalone scheme, which only worsen with distance. Standalone nodes send messages at 0.333\,Hz  and close neighbors listen to them at closer to 0.2\,Hz. Nevertheless, at short distances, \acp{IPG} stay considerably low for the implicit clustering scheme, and it is until the 400--500~\,m segment that losses are significant (i.e., we are blind to a \ac{VRU} or a cluster for more than 8\,s when we try to send messages every 4\,s). The shortcomings of the standalone scheme have implications for ETSI clustering, since it would be hard for the system to converge into clustering starting from an all-standalone point if individual messages have so low of a success rate. Further work is being performed to confirm these implications.

\begin{figure}
    \centering
    \includegraphics[width=1\linewidth]{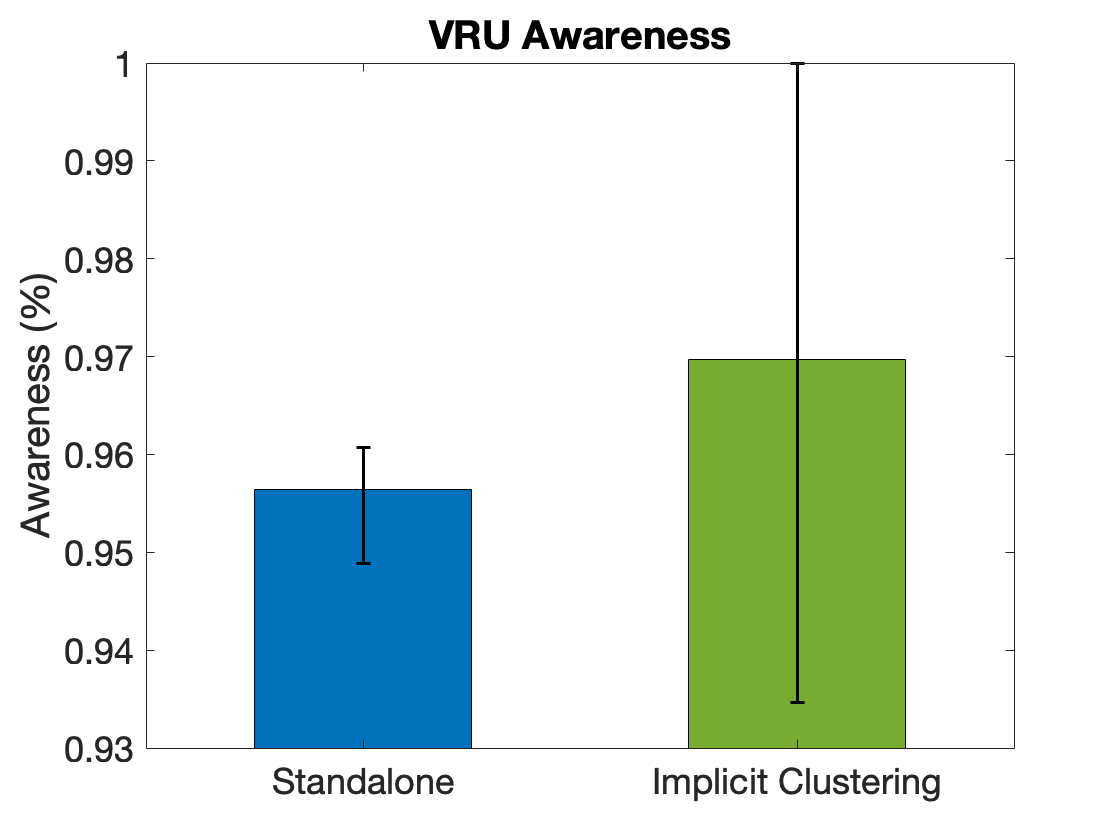}
    \caption{VRU awareness ratio with VRU and vehicular traffic}
    \label{fig:awareness_veh}
\end{figure}

However, it could be considered trivial to compare a scheme that reduces network contention and not expect it to improve network performance. Therefore, a comparison for safety metrics shows the real effect of reducing the number of messages entering the medium on VRU awareness. Fig.~\ref{fig:awareness_veh} shows that safety metrics stay also at acceptable levels when vehicular traffic enters the scenario. The clustered approach has a better average but more variability. We aggregate both VRU-to-VRU and VRU-to-Vehicle perception since there are types of VRUs that are closer to \textit{small vehicles}. These interactions between types of \acp{VRU} are also interesting to watch since the ETSI ITS framework has pedestrians, bikes, and even animals sending VAMs. Therefore, studying these interactions in the future is important for: 1) safety when avoiding, e.g., crashes between a bicycle and a pedestrian; and 2) network performance, e.g., to be able to cluster VRUs implicitly or explicitly by type and dynamics, or that more dynamic VRUs (e.g., \acp{EPAC}) do not overwhelm pedestrian messages (e.g., an \ac{EPAC} might be able to support a more powerful \ac{ITS-S}).

These results are also important for the clustering scheme proposed by ETSI. Starting from the point where leaders are required to send more messages (i.e., every 2\,s for keeping the cluster alive, plus those with a limit of 0.5\,s to notify members of a successful joining operation), and success rates being as low as 60\% at short distances, it is safe to assume that some messages for cluster operations will not be received and this has implications on safety metrics. Further work is required to measure these phenomena and effects.

\section{Discussion and Related Work}
\label{sec:discussion}

Previous state-of-the-art works on this subject \cite{silasCluster, xhoxhiClusteringCPM} have focused on implementing the clustering functionality precisely as defined in the ETSI VBS \cite{etsi-va} while providing some recommendations for potential improvements. More specifically, in \cite{silasCluster} the findings indicate that while clustering VRUs does optimize spectrum resource allocation, the channel load can still be improved by providing the cluster leader with the information about a VRU’s intention to join a cluster. In this work, we proposed an non-negotiated clustering technique that avoids continuous broadcasting of cluster-VAMs (which are larger than regular VAMs) thus bypassing the issue of needing to communicate the intention to the cluster leader and reducing channel load. We have shown, that our algorithm not only directly improves the average IPG which means the information from neighboring vehicles is more up-to-date but, it also greatly reduces the difference between the average IPG and IGG values compared to the standalone communication mode, improving network reliability. In addition we have measured the performance of our proposed scheme with and without network traffic from vehicles (CAMs) and found that the performance metrics remain stable even in the presence of higher channel utilization levels which demonstrates the scalability of our proposed solution.

\section{Conclusions and Future Work}
\label{sec:conclusion}
We presented an implicit clustering scheme for ETSI VAMs where cluster operations and maintenance not performed explicitly. Our clustering scheme deals intrinsically with problems related to network phenomena, namely collisions, that make other road users \textit{blind} to a cluster, and cluster members enter uncertainty about whether or not they are still part of a cluster. Furthermore, our scheme does not require signaling for joining or leaving a cluster, avoiding the use of additional data structures or even unnecessary messages. 

Our results show that implicit clustering keeps VRU awareness levels while surpassing network metrics. This paves the way for other works in progress, e.g., in the field of VRU intention sharing, where we are working on keeping VAM sizes constant while sharing enough information to convey a VRU's future location~\cite{tesis_johan}. Our future work includes the mixing of clustering and intention sharing, i.e., include a VRUs future position as a condition for clustering, thus, intrinsically keeping homogeneity.

Finally, a more extensive analysis of our clustering scheme is due. First, by including more types of VRUs (e.g., cyclists, scooters, and other Type I-II~\cite{etsi-vam-use} VRUs) and several road user densities. Furthermore, a comparison with other clustering schemes proposed in the literature is also a future step.



\bibliographystyle{ieeetr}
\bibliography{references}

\end{document}